\documentclass{sig-alternate}

\newcommand{\utt}[1]{\unskip\hspace{.16667em plus .08333em}\texttt{#1}}
\newcommand{\uttt}[1]{\texttt{#1}}

\usepackage[font=bf,skip=\baselineskip]{caption}
\DeclareCaptionType{copyrightbox}
\usepackage{subcaption}

\usepackage{placeins}
\usepackage{url}

\begin{document}
%
\toappear{\copyright~Scott A.~Hale, Taha Yasseri, Josh Cowls, Eric T.~Meyer, Ralph Schroeder, and Helen Margetts, 2014. This is the authors' version of the work. 
It is posted here for your personal use. Not for redistribution. The definitive version is published in WebSci~'14, \url{http://dx.doi.org/10.1145/2615569.2615691}.}

\newfont{\mycrnotice}{ptmr8t at 7pt}
\newfont{\myconfname}{ptmri8t at 7pt}
\let\crnotice\mycrnotice%
\let\confname\myconfname%


\title{Mapping the UK Webspace:\\
Fifteen Years of British Universities on the Web}

%
%
%
%
%

\numberofauthors{6} 
%

\newcommand{\oii}{\affaddr{Oxford Internet Institute}\\%
	\affaddr{University of Oxford}\\%
	\affaddr{1 St Giles, Oxford UK}}

\author{
\alignauthor
Scott A. Hale\\
		\oii\\
       \email{scott.hale@oii.ox.ac.uk}
\alignauthor
Taha Yasseri\\
       \oii\\
       \email{taha.yasseri@\ldots}
\alignauthor Josh Cowls\\
       \oii\\
       \email{josh.cowls@\ldots}
\and  
\alignauthor Eric T. Meyer\\
       \oii\\
       \email{eric.meyer@\ldots}
\alignauthor Ralph Schroeder\\
		\oii\\
       \email{ralph.schroeder@\ldots}
\alignauthor Helen Margetts\\
		\oii\\
       \email{helen.margetts@\ldots}
}
\date{23 February 2014}

\maketitle
\begin{abstract}
This paper maps the national UK web presence on the basis of an analysis of the \utt{.uk} domain from 1996 to 2010. It reviews previous attempts to use web
archives to understand national web domains and describes the dataset. Next, it presents an analysis of the \utt{.uk} domain, including the overall number
of links in the archive and changes in the link density of different second-level domains over time. We then explore changes over time within a particular 
second-level domain, the academic subdomain \utt{.ac.uk}, and compare linking practices with variables, including institutional affiliation, league table
ranking, and geographic location. We do not detect institutional affiliation affecting linking practices and find only partial evidence of league table 
ranking affecting network centrality, but find a clear inverse relationship between the density of links and the geographical distance between universities. 
This echoes prior findings regarding offline academic activity, which allows us to argue that real-world factors like 
geography continue to shape academic relationships even in the Internet age. We conclude with directions for future uses of web archive resources 
in this emerging area of research.
\end{abstract}




\section{Introduction}
The World Wide Web is enormous and is in constant flux, with more web content lost to time than is currently accessible via the live web. 
The growing body of archived web material available to researchers is thus potentially immensely valuable as a record of important aspects 
of modern society, but there have previously been few tools available to facilitate research using archived web materials \cite{dougherty2014}. 
Nevertheless, with the development of new tools and techniques such as those used in this paper, the use of web archives both to understand the 
history of the web itself as well as to shed light on broader changes in society is emerging as a promising research area \cite{dougherty2010}. 
The web is likely to provide insight into social changes just as other historical artifacts, such as newspapers and books, have done for scholars
interested in the pre-digital world. As the web becomes increasingly embedded in all spheres of everyday life and the number of webpages continues to grow, there is a 
compelling case to be made for examining changes in both the structure and content of the web. However, while interfaces such as the Wayback 
Machine\footnote{\url{http://web.archive.org/}} allow access to individual webpages one at a time, there have been relatively few attempts 
to work with large collections of web archive data using computational approaches across the corpus. This paper provides a longitudinal 
analysis of the UK national web domain, \utt{.uk}, and the academic second-level domain, \utt{.ac.uk}, in order to show the benefits and challenges of this type of analysis.

\section{Background}

\subsection{Archiving national web domains}
National web domains represent one approach to web archive analysis for researchers seeking an overview of a single country's web presence. 
A particular national web domain offers the potential of both diversity and completeness in its coverage \cite{baeza2007}, although there are 
limitations in terms of generalizability beyond the country in question and frequently in terms of the completeness of the analysis based on 
technical factors (see below). At the same time, however, limiting the focus to a single country also has the potential to introduce fewer 
contextual differences (such as language, Internet penetration rates, broadband penetration rates, political openness, economic differences, 
and so forth), and thus is a sound strategy for demonstrating the potential of this type of analysis, which has not previously been done.

Research in this area is at an early stage, and there are conceptual challenges associated with analyzing national web domains.
The content and structure of country-code top-level domains (ccTLDs) such as \utt{.uk} for the United Kingdom and \utt{.fr} for France are governed more by
traditions than rules \cite{masanes2006}, complicating efforts to reach a comprehensive definition of what they represent. 
Br\"ugger \cite{brugger2014} discusses the difficulty, for example, of deciding how national presences should be delimited. In the case presented here,
the domain name \utt{.uk} is used, but this does not cover all the webpages originating in the United Kingdom as several British companies, organizations,
and individuals operate domains in generic top-level domains (\uttt{.com}, \utt{.org}, etc.) or elsewhere. Moreover webpages ending with \utt{.uk} are also used for
websites which arguably belong to a different country, as when multinational companies headquartered outside the UK have affiliates within the UK with 
a \utt{.uk} address. Finally, it might be
contended that not only webpages with a \utt{.uk} address be examined, but also those that link to and from these webpages. However, for the purposes of
this research, these limitations can mostly be noted for future research and do not seriously limit the ability to understand the broad patterns
within the UK national web presence.

Another issue that must be decided when undertaking analysis of web domains is the appropriate level of detail. This includes the temporal
resolution to use for analysis (since while the web is constantly changing, the number of snapshots available in Internet Archive data vary over time 
based on the crawl settings in place when the data were gathered) and what level of detail to extract from webpages (i.e., determining the appropriate
level of resolution of page content, link information, page metadata, and so forth).
Previous research on the \utt{.uk} ccTLD has examined monthly snapshots over a one year period finding page-level hyperlinks change frequently month to month \cite{bordino2008}.
As Br\"ugger \cite{brugger2013} notes, there are several reasons
why archived websites are different from other archived material in respect to these details: choices must be made not just about what to capture but 
also technical issues about what can be archived and how the archiving process itself shapes the later availability of the archived materials.

\subsection{UK web domain}
For the \utt{.uk} domain that will be examined here, the source of the data is the archive files of the UK domain that were obtained from the Internet
Archive by the British Library with the specific purpose of creating the basis of a national archive of the web in the UK. This data is
currently being expanded via ongoing web archiving activities being performed by the British Library under the terms of the 2003 UK legal deposit law,%
\footnote{Legal Deposit Libraries Act 2003, \url{http://www.legislation.gov.uk/ukpga/2003/28/contents}}
which was implemented via new regulations that went into effect in April 2013.%
\footnote{The Legal Deposit Libraries (Non-Print Works) Regulations, \url{http://www.legislation.gov.uk/uksi/2013/777/contents/made}}

The \utt{.uk} country-code top-level domain is managed by the Internet registrar Nominet.\footnote{\url{http://www.nominet.org.uk/}}
Below the \utt{.uk} top-level domain are several second-level domains (SLDs), the largest of which are \utt{.co.uk} (commercial enterprises), \utt{.org.uk} (non-commercial organizations),
\utt{.gov.uk} (government bodies), and \utt{.ac.uk} (academic establishments).\footnote{\url{http://www.nominet.org.uk/uk-domain-names/about-domain-names/uk-domain-subdomains/second-level-domains}}
This paper examines the data aggregated to the level of third-level domains such as \uttt{nominet.org.uk} (Nominet), \uttt{fco.gov.uk} (the Foreign and Commonwealth Office of the UK government), or \uttt{ox.ac.uk} (the University of Oxford).

In the case of web archives (or indeed of other archived material which takes the approach of archiving all that can be archived, without 
a particular topic in mind), it is not scholarly interest that sets the agenda, but rather the goal of the
archiving institution. This means that the scope of the archived material and the level of detail available, as with other historical materials, 
is a function of the archiving processes used to gather and store the data. Thus, unlike web archive research done on the live web using
researcher-implemented data collection mechanisms [e.g., \cite*{escher2006,foot2006}], for the purpose of this study the dataset itself should be seen as a given. However, it can be mentioned that the Internet Archive's data comprises the most comprehensive archive of the web available \cite{ainsworth2011}.

\section{Data}
\subsection{Data preparation}
The data for this study originally comes from the Internet Archive, which began crawling pages from all domains in 1996 \cite{kahle1997}. 
Copies of the approximately 30 terabytes of compressed data relating to the \utt{.uk} country-code top-level domain (ccTLD) was provided to the
British Library and forms the ``JISC UK Web Domain Dataset.''\footnote{\url{http://data.webarchive.org.uk/opendata/ukwa.ds.2/}}

Hale et al.~\cite{jisc_intro} cleaned the data by removing error pages (e.g.~404 Not Found pages) as well as pages not within the \utt{.uk} ccTLD.
They produced a plain-text list of all page urls remaining in the collection and the date and times they were crawled, and an additional 
plain-text list of all outgoing hyperlinks starting from pages within the dataset.

For this study, we started with this list of hyperlinks and filtered it to only include links between different third-level domains. We further grouped pages crawled at similar times (within 1,000 seconds) together and assigned the hyperlink pair a weight based on the number of hyperlinks between the two third-level domains in that time period. For each year, we take the crawl with the largest number of captured hyperlinks between any two domains. We also formed one list of all third-level domains present in the dataset each year and the number of pages crawled within each third-level domain. We loaded these lists into Apache Hive for further analysis.

\newpage
\subsection{Data analysis}
In what follows, we undertake a longitudinal network analysis, charting the \utt{.uk} domain and its core second-level domains over time.
As Br\"ugger \cite{brugger2013} points out, this type of analysis is not concerned with who produced what, nor with how the web content
was used, but rather with what was created and thus ``the web which is''---or rather was---``actually available to users.''

First, we present an overall longitudinal view of the second-level domains within the \utt{.uk} domain. We investigate the growth of the entire domain between 1996 and 2010, broken down into its four largest constituent parts, \utt{.co.uk}, \utt{.org.uk}, \utt{.gov.uk}, and \utt{.ac.uk}.
Analysis of these SLDs allows us to investigate the role of different sectors of British society in the growth of the UK web presence.

The second section looks at the linking practices between these SLDs. It asks about the internal link density of each SLD, and analyses how they 
interact with each other: whether, for example, there are more links between certain subdomains, and whether linking is reciprocal between domains or imbalanced.

The third and final section of the findings takes a closer look at the SLD \utt{.ac.uk}. It builds on earlier longitudinal analyses of academic webpages,
which have investigated, for example, the stability of outlinks \cite{payne2008,thelwall2003}. Our findings update earlier
studies by extending the period of analysis to the end of 2010 and assessing the effects of new variables, including institutional affiliation, 
league table ranking, and geographic location on link practices between different universities.

\section{Results}
\subsection{Overview of .uk}
\begin{figure}
\includegraphics[width=\columnwidth]{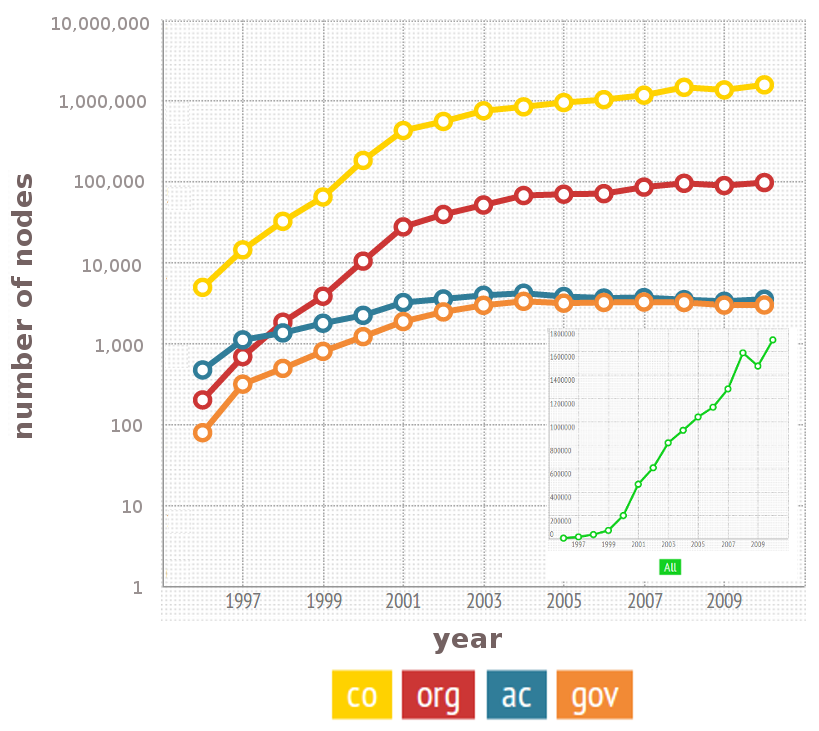}
\caption{Number of nodes (third-level domains) within each second-level domain over time. The inset shows the sum over all second-level domains.}
\label{fig:overall_growth}
\end{figure}

Figure \ref{fig:overall_growth} displays the overall growth of the \utt{.uk} ccTLD, showing the total number of nodes (on a logarithmic scale) within
each SLD from 1996 to 2010. It also shows the size of the entire \utt{.uk} domain space (on a linear scale). There is a clear change in the trend of the growth around 2001 for \utt{.co.uk} and \utt{.org.uk}.
Furthermore, \utt{.ac.uk} and \utt{.gov.uk} seem to almost stabilize in size at around the same time.

\begin{figure}
	\includegraphics[width=.9\columnwidth]{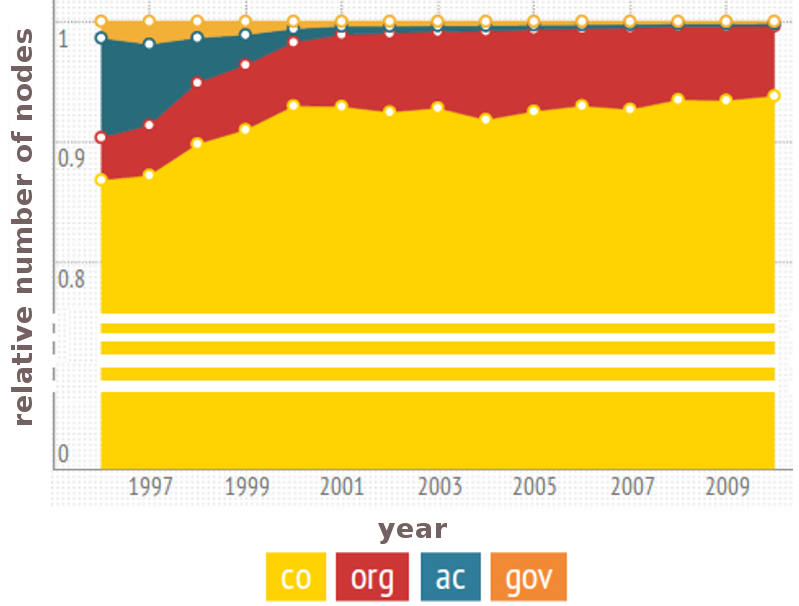}
	\caption{Relative size of second-level domains in the \utt{.uk} top-level domain over time.}
	\label{fig:sld_relative}
\end{figure}

Figure \ref{fig:sld_relative} shows the relative size of the second-level domains \utt{.co.uk}, \utt{.org.uk}, \utt{.ac.uk}, and \utt{.gov.uk} across the fifteen year period,
standardized as each SLD's proportion of the total nodes (i.e., domains/websites, not webpages) in the collection in each year. While these are not the only
second-level domains in use within the \utt{.uk} domain, they are the four largest in terms of number of nodes across the whole period.


As Figure \ref{fig:sld_relative} shows, \utt{.co.uk} is the predominant second-level domain throughout the entire period, with \utt{.co.uk} sites never accounting
for less than 85\% of the total. However, also apparent is the large proportion of governmental and, especially, academic sites in the early recorded 
history of the UK web. This is consistent with the role that universities played in the early establishment, adoption, and development of
the web \cite{leiner2009}. Over time, however, this early presence was greatly overshadowed in terms of absolute numbers of nodes
when compared to the continued growth of the \utt{.co.uk} and \utt{.org.uk} domains.

\newpage
\subsection{Link density among and between second-level domains}
\begin{figure}
	\includegraphics[width=\columnwidth]{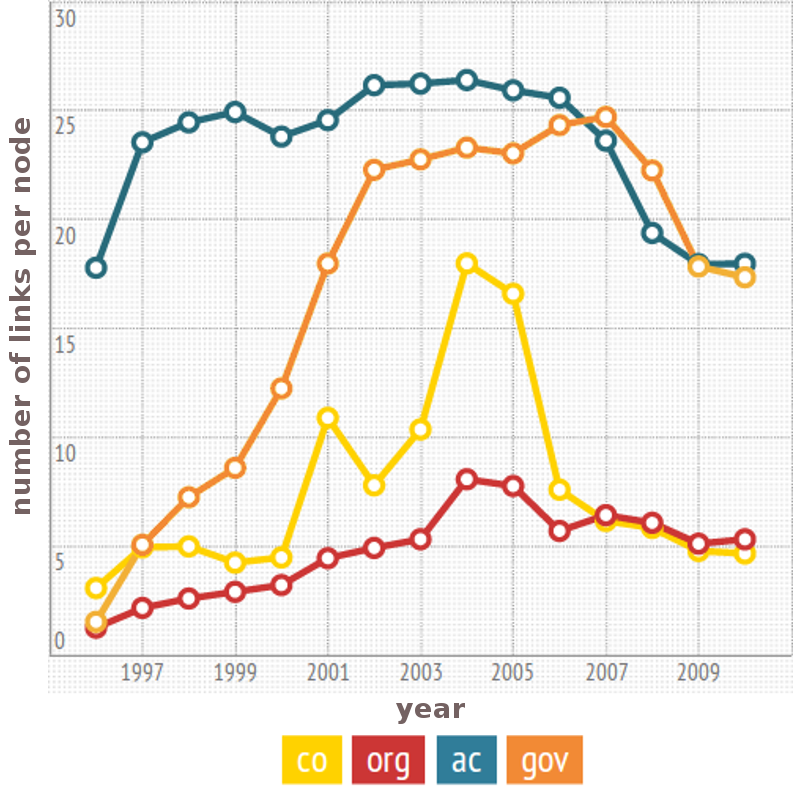}
	\caption{Number of within-SLD links per node in four \utt{.uk} SLDs, 1996--2010.}
	\label{fig:links_per_node}
\end{figure}

Up to this point the analysis has drawn only on node data; that is, the number of websites making up each domain. However, link analysis can offer 
insight into how well integrated each SLD is with itself and with other domains.
A link from one site to another has been used as an indicator of awareness between blogs \cite{hale2012} and recognition between academic sites \cite{thelwall2003}.
Figure \ref{fig:links_per_node} shows, for each subdomain, how many links there are for every node over time,
where a fluctuating relationship between the number of nodes and links to other nodes for each second-level 
domain is visible. Over the whole period, the \utt{.ac.uk} academic SLD and, from 1997 onwards, the \utt{.gov.uk} governmental SLD are the most internally dense SLDs. This 
observation may reflect the fact that registration for the \utt{.ac.uk} and \utt{.gov.uk} subdomains is restricted, whereas \utt{.org.uk} and \utt{.co.uk} sites can be registered 
easily by any party. In addition, the \utt{.ac.uk} and \utt{.gov.uk} subdomains are likely constituted by a narrower and more cohesive set of institutions, creating,
on average, a stronger basis for linking within the SLDs. Furthermore, there is likely more competition and thus less reason to link within
the \utt{.co.uk} commercial subdomain compared to \utt{.ac.uk} or \utt{.gov.uk}.
Higher link density within the \utt{.org} and \utt{.gov} domains in comparison to the \utt{.com} domain has previously been observed in a smaller scale, topical study about climate change \cite{rogers2000}.

Also of note is the general rise of links in the middle of the period, particularly in the substantial \utt{.co.uk} subdomain. This peaks sharply in 
2004 before falling sharply back to around pre-2001 levels by 2009. This trend has no easy explanation, suggesting that further research is 
required to explain this pattern. Possible explanations include the norm of including lists of links on webpages such as blogs fell out of 
favor in the middle of this period or that more websites increasingly linked outside of the \utt{.uk} ccTLD. 

\begin{figure*}
	\begin{center}
	\begin{subfigure}{0.5\textwidth}
		\begin{center}
			\includegraphics[width=\textwidth,height=7cm,keepaspectratio]{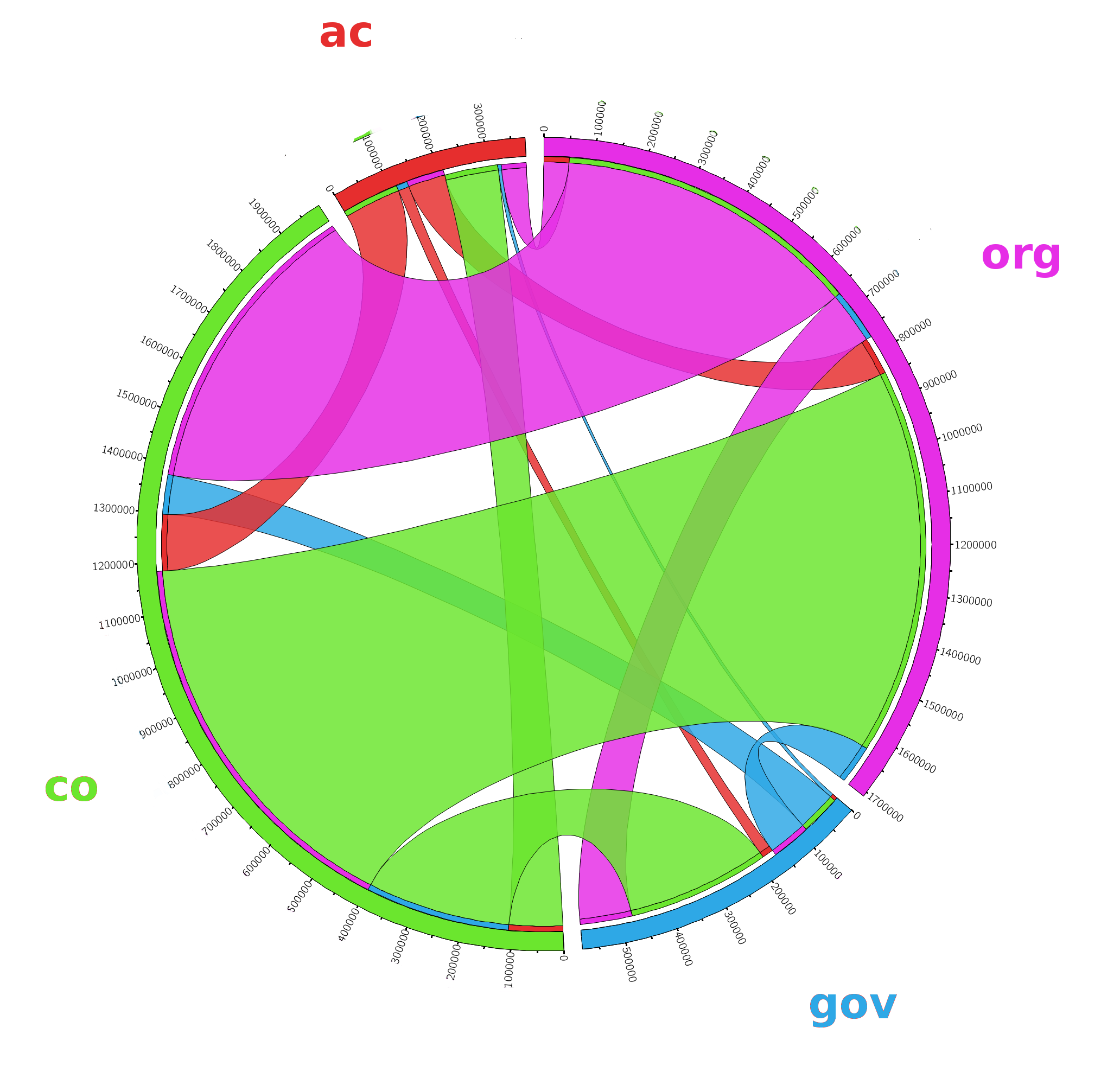}
			\caption{}
			\label{sld_absolute}
		\end{center}
	\end{subfigure}%
	\begin{subfigure}{0.5\textwidth}
		\begin{center}
			\includegraphics[width=\textwidth,height=7cm,keepaspectratio]{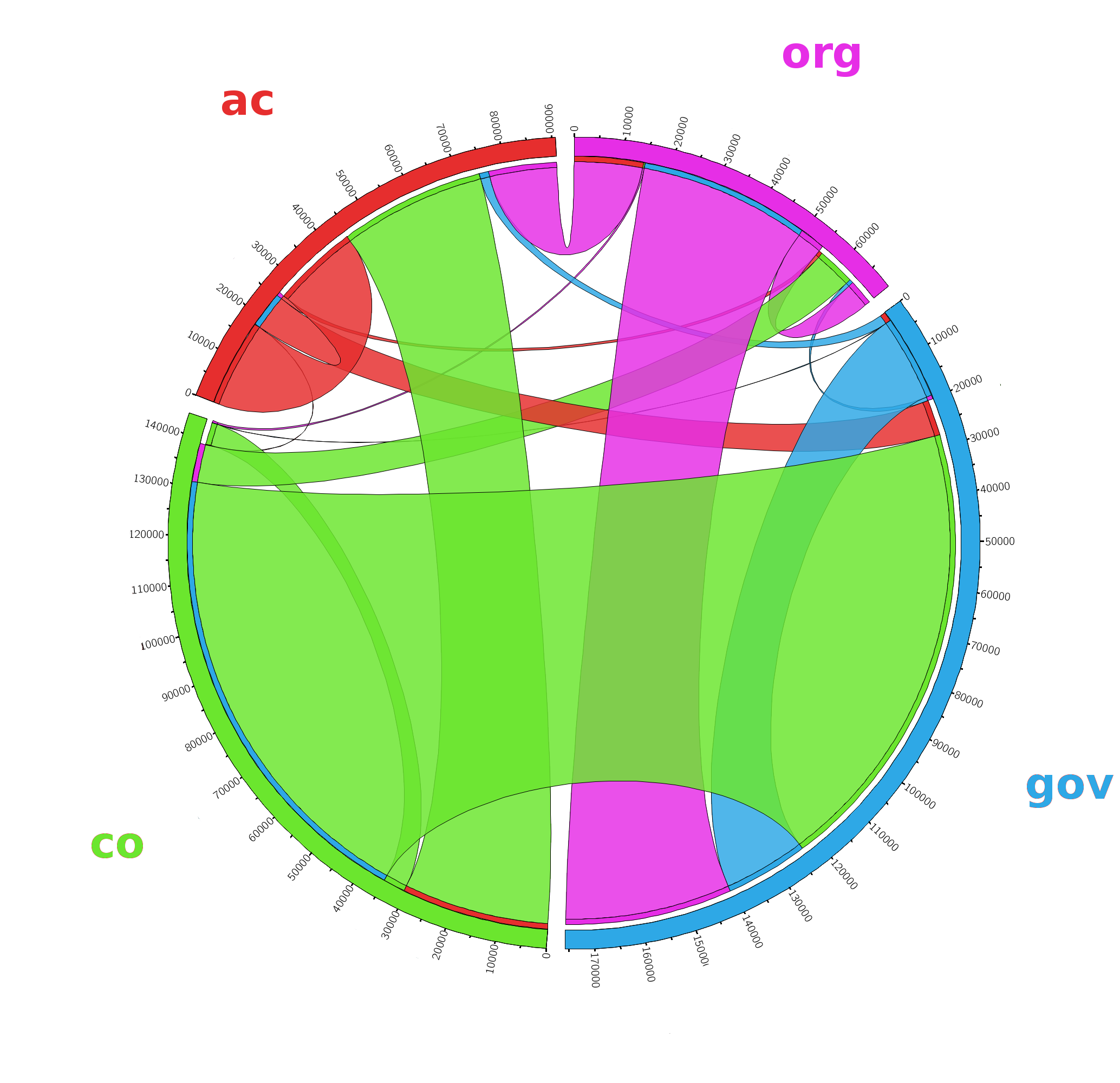}
			\caption{}
			\label{sld_normalized}
		\end{center}
	\end{subfigure}
	\caption{Links between four second-level domains. Panel \textit{a} shows the absolute number of links between different SLDs (self-loops are excluded), and panel \textit{b} shows the relative number of links normalized by the size of target subdomain.}
	\label{fig:sld_interlink}
	\end{center}
\end{figure*}

Not only can web domain data tell us how well integrated an SLD is internally, but we can also investigate how well SLDs are connected 
to each other. Figures \ref{fig:sld_interlink}a and \ref{fig:sld_interlink}b show the quantity of links between SLDs for 2010,
the last year in the dataset, where the size of an arc relates to the volume of links from one SLD to another. The color of each arc relates
to links sent in one direction, from the host SLD outwards. For example green arcs show links from the \utt{.co.uk} domain to others.
Figure \ref{fig:sld_interlink}a shows the absolute volume of links, while the size of the arcs in Figure \ref{fig:sld_interlink}b are normalized in 
relation to the number of nodes in the target subdomain. (Note that Figure \ref{fig:sld_interlink}a does not display links within a single SLD, as
the volume of links between \utt{.co.uk} sites dwarfs all other relationships. As Figure \ref{fig:sld_interlink}b controls for the number of nodes in each SLD, the adjusted \utt{.co.uk} arc is
much smaller and links within a single SLD are therefore included.)

Figure \ref{fig:sld_interlink}a shows that the largest volume of links between SLDs in 2010 flowed from \utt{.co.uk} sites to \utt{.org.uk} sites, 
and this relationship is fairly reciprocal, with \utt{.org.uk} sites sending almost as many links back. Links between other domains are much 
lower in terms of absolute volume. When controlling for the size of the target subdomain, however, the picture changes somewhat. 
As Figure \ref{fig:sld_relative} above showed, by 2010 the number of nodes in the \utt{.org.uk} subdomain far outweighed those in the \utt{.ac.uk} 
and \utt{.gov.uk} subdomains. Figure \ref{fig:sld_interlink}b, adjusting for this, shows that the \utt{.gov.uk} and, to a lesser extent, the \utt{.ac.uk} 
subdomains punch above their weight, receiving proportionally more links from \utt{.co.uk} and \utt{.org.uk} sites. Once again, the more restrictive
registration policies for these SLDs may be a factor here, driving up the average quality and `linkworthiness' of sites in these 
subdomains as compared to \utt{.co.uk} and \utt{.org.uk} sites
although it may also be related to other factors such as the comparative homogeneity of these SLDs, the perception of objectivity or balance on academic or government websites as opposed to sites oriented towards sales or persuasion, or even the international standing of many UK universities, although understanding these factors would require further investigation.

For the \utt{.gov.uk} subdomain, the finding that sites link out less than they are linked to suggests a lack of `outward-lookingness,' 
compared to the other sectors. In contrast, Escher et al.~\cite{escher2006} found the UK Foreign and Commonwealth Office to be relatively 
more outward-looking than its equivalents in Australia and the US. However, foreign offices, with their outward facing role could easily be an
exception to more general government-wide propensity not to link out.

In addition, it is worth noting the relatively heavy proportion of links within the \utt{.ac.uk} SLD shown in Figure \ref{sld_normalized}. This propensity of academic institutions to link heavily to other academic institutions (more so than the other domains) reflects (taking a positive view) a strong network among academic institutions, but also potentially (taking a negative view) a tendency towards inward-looking, within-domain links. We examine these links in more depth in the next section.


\subsection{The case of the academic subdomain}
At this stage we turn our attention to one particular subdomain, the \utt{.ac.uk} academic subdomain of the UK web. To be eligible for a third-level domain within \utt{.ac.uk}, an organization must have a permanent physical presence in the UK and either have the majority of its activities publicly funded by UK government funding bodies or be a Learned Society. In addition, the organization must satisfy at least one of the following criteria: the organization must provide tertiary-level education with central government funding, conduct publicly funded academic research, have a primary purpose of supporting tertiary-level educational establishments, or have the status of a Learned Society (``a society that exists to promote an academic discipline or group of disciplines'').\footnote{\url{https://community.ja.net/library/janet-services-documentation/eligibility-guidelines}}

The academy was at the 
forefront of the development of the web, and, as Figure \ref{fig:sld_relative} shows, \utt{.ac.uk} sites constituted a sizeable minority 
of \utt{.uk} sites in 1996. Over time, this proportion waned, even as more British universities established a substantial web presence. 
In this subsection we use the longitudinal data collected to examine the relationship between universities' linking practices and 
three variables: institutional affiliation, league table ranking, and geographic location. Our hypothesis in doing so was that higher 
status academic institutions would be more strongly linked to than lower status institutions and would also be more strongly 
interconnected with their peer institutions.

For the analysis, we built a list of the 121 universities listed in the most recent Sunday Times University Guide.%
\footnote{\url{http://www.thesundaytimes.co.uk/sto/University_Guide/}}
Each of these universities has a website, all of which use the \utt{.ac.uk} suffix. We obtained the third-level domain (e.g., \utt{ox.ac.uk}) for each.
Further data collection as necessary is described in the respective subsections that follow.

\subsubsection{Group affiliation}
\begin{figure}
	\includegraphics[width=\columnwidth]{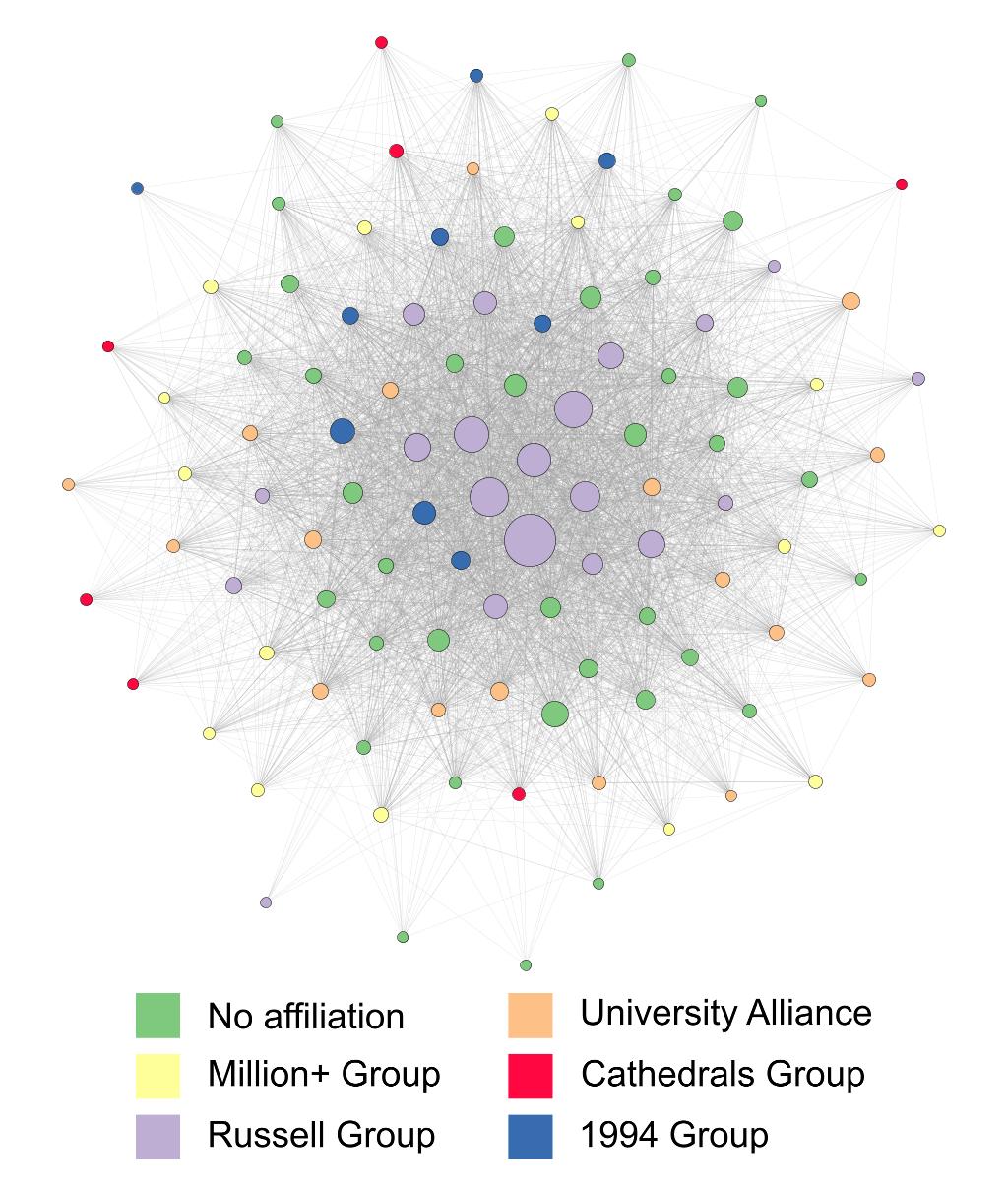}
	\caption{Network diagram of hyperlinks between universities. Different colors indicate different university affiliations.}
	\label{fig:ac_network}
\end{figure}

Many British universities belong to associations, formed to represent their interests and facilitate collaboration. The groups are neither mutually 
exclusive nor exhaustive, meaning that universities can belong to none, one, or more than one group, but for practical and political reasons most 
universities belong to only one. We collected data on the memberships of five groups,
the Russell Group,\footnote{\url{http://www.russellgroup.ac.uk/our-universities/}}
the 1994 Group,\footnote{\url{http://www.timeshighereducation.co.uk/news/was-1994-groups-demise-triggered-by-relaunch-delays/2008999.article}}
the University Alliance,\footnote{\url{http://www.unialliance.ac.uk/member/}}
the Million\texttt{+} Group,\footnote{\url{http://www.millionplus.ac.uk/who-we-are/our-affiliates/}}
and the Cathedrals Group.\footnote{\url{http://cathedralsgroup.org.uk/Members.aspx}}

The best known of these is perhaps the Russell Group, formed in 1994 and now constituted of 24 members. The 1994 Group, which represented smaller 
research institutions, was formed in response to the Russell Group, but disbanded in 2013; given the time frame of the dataset we include the 11 
final members of the group in our analysis. Of the final three groups, the University Alliance is formed of 22 business-oriented UK universities, 
the Million\texttt{+} Group is made up of 17 mostly `new' (post-1992) institutions, and the Cathedrals Group is made up of 16 universities 
originally instituted as church-led teacher training colleges. The stated purposes of these groups differ somewhat, but each are constituted 
broadly to serve the research interests of their members.

In comparing group membership to the density of links between different universities, we sought to discover whether academic affiliation was 
associated with the density of links between institutions. To do this, we performed a network analysis, investigating whether the universities 
clustered on the basis of group affiliation. Figure \ref{fig:ac_network} shows a network diagram, with different affiliations marked by different colors.

To the naked eye, Figure \ref{fig:ac_network} shows no discernible clustering on the basis of group affiliation, and network analysis bears this out. 
The division of the network by affiliations has a modularity score \cite{newman2006} of $-0.003$, indicating that the division of the network into clusters based on university
affiliation is no better than dividing the network into five random clusters. On an individual basis, only one group, the Russell Group, has many internal
links and comparatively fewer links to institutions outside the group. It is the most strongly connected group with an internal hyperlink density of 0.71. 
The Russell Group, which includes 24 of the leading international UK universities with some of the highest levels of research funding, arguably represents 
most if not all of the elite universities in the UK. It contains 9 of the 10 top-ranked UK universities, including both Oxford and Cambridge. 
That these universities are more strongly linked to each other is likely related at least in part to
their active research cultures, with many collaborations between researchers at these top institutions. The lack of strong web connections in the other associations, 
however, suggests that while these institutions may or may not have strong connections among their members by other measures, there is no evidence that universities
strongly link to the websites of institutions with which they share group affiliation over other institutions.

\subsubsection{League table ranking}
\begin{figure}
	\begin{center}
		\includegraphics[width=\columnwidth]{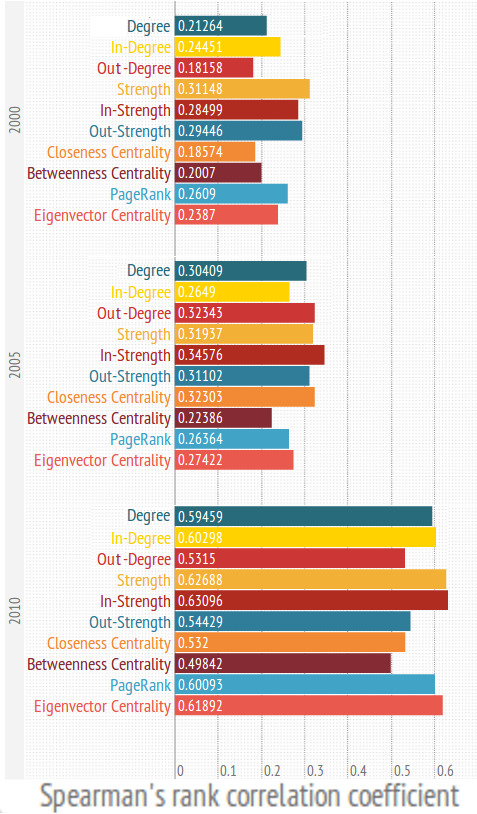}
		\caption{Spearman's rank correlation coefficients between university league table rankings and ten different network centrality measures for three years.}
		\label{fig:spearman_centrality}
	\end{center}
\end{figure}

University league tables are an important if imperfect indicator of a university's prominence. Modern league tables incorporate a whole range of measures, 
including factors related to teaching, research, and student satisfaction. As such, we wanted to investigate whether a university's league table ranking is 
associated with its web presence, and whether the relationship has changed over time, in terms of both increasing adoption and development of an institution's 
web presence and its changes in league table ranking over time. For this analysis, we collected the rankings of British universities published in 
The Times Good University Guide for three years, 2000, 2005 and 2010, and compared these rankings with data from crawls conducted in the same 
three years.

In conducting the analysis, we used ten common measures of network centrality for each of the three different years to gauge the relationship between
each university's league ranking and its position in the network of hyperlinks flowing between university third-level domains. We then produced lists ranking
the universities for each year by each centrality measure and computed Spearman's rank correlation coefficient for each centrality ranking and league table ranking combination.
These correlation coefficients are shown in Figure \ref{fig:spearman_centrality}.


\begin{figure}
	\begin{center}
		\includegraphics[width=0.4\textwidth]{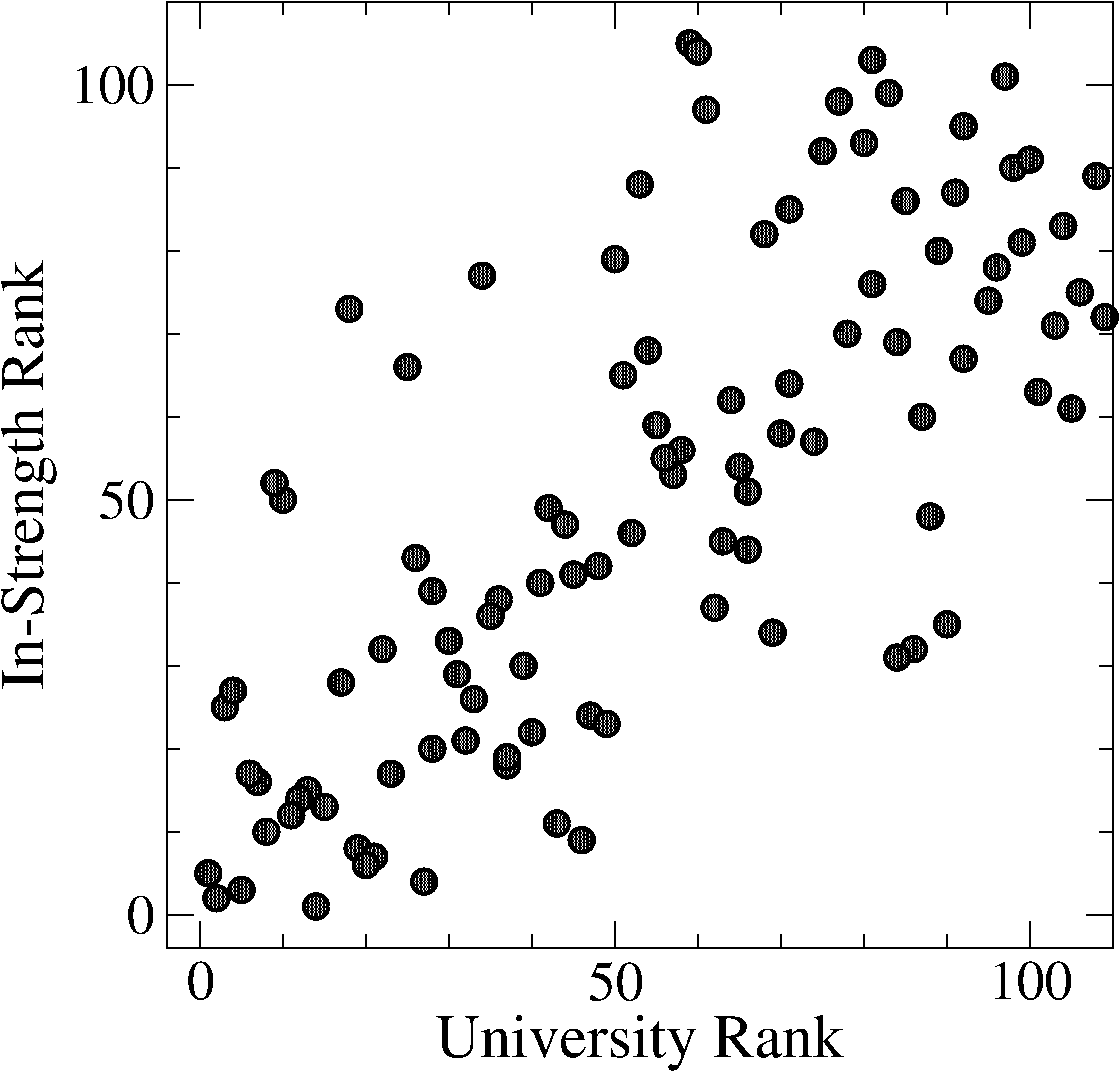}
		\caption{University in-strength rankings compared to university league table rankings for 2010. Spearman's rank correlation is 0.63.}
		\label{fig:strength_rank_scatter}
	\end{center}
\end{figure}

For most measures of centrality used, a pattern emerges: the data for 2010 shows the strongest correlation between league table ranking and centrality, 
while the relationship is less evident for 2000 and 2005. The most strongly correlated correlation measure is in-strength, a sum of all the hyperlinks
linking to a given web domain. This measure uses the weight of each edge, which corresponds to the number of hyperlinks between any two third-level domains. This differs from in-degree which measures the number of other domains that link to a given web domain.
Figure \ref{fig:strength_rank_scatter} shows the fairly strong correlation between universities' league table rankings and their 
network positions as measured by in-strength. What Figure \ref{fig:spearman_centrality} and Figure \ref{fig:strength_rank_scatter}
suggest is two-fold: first, that university prominence, as measured by league table position, is an increasingly stronger predictor of 
the number of links to that institution over the 2000--2010 period. Whether this is an example of the Matthew Effect (``the rich get 
richer'') \cite{merton1968} whereby highly prominent institutions become well-linked institutions largely as a result of their prominence 
(and conversely, marginal institutions become more marginalized as a result of their lack of prominence), or whether there is another 
independent factor at play here cannot be determined from these data. However, the second conclusion is clear:
the hyperlink patterns within the UK academic subdomain support the notion that the web does not inherently challenge existing power structures.
Instead, the saturation of the \utt{.ac.uk} subdomain, in terms of the presence of essentially all possible academic institutions by 2003
(Figure \ref{fig:overall_growth}), has resulted by 2010 in a subdomain in which network centrality closely mirrors prominence as measured by league tables.


\subsubsection{Role of geography}
Finally, we investigated whether any association exists between the geographic proximity of British universities and the density of hyperlinks 
between them. This analysis builds upon work by Pan et al.~\cite{pan2012} who found, at a global scale, that rates of academic citations and 
collaborations between two cities diminish as the distance between them increases, following gravity laws. We conduct a similar analysis, 
replacing citations and collaborations with hyperlinks collected in the web domain data. 

We collected geographic coordinates for each of the British universities in the list using a simple Google Maps search. Universities can be 
spatially complex, sometimes having multiple campuses and satellite sites; so, some discretion was occasionally required in identifying the center of each university.

\begin{figure*}
	\begin{center}
	\includegraphics[width=0.8\textwidth]{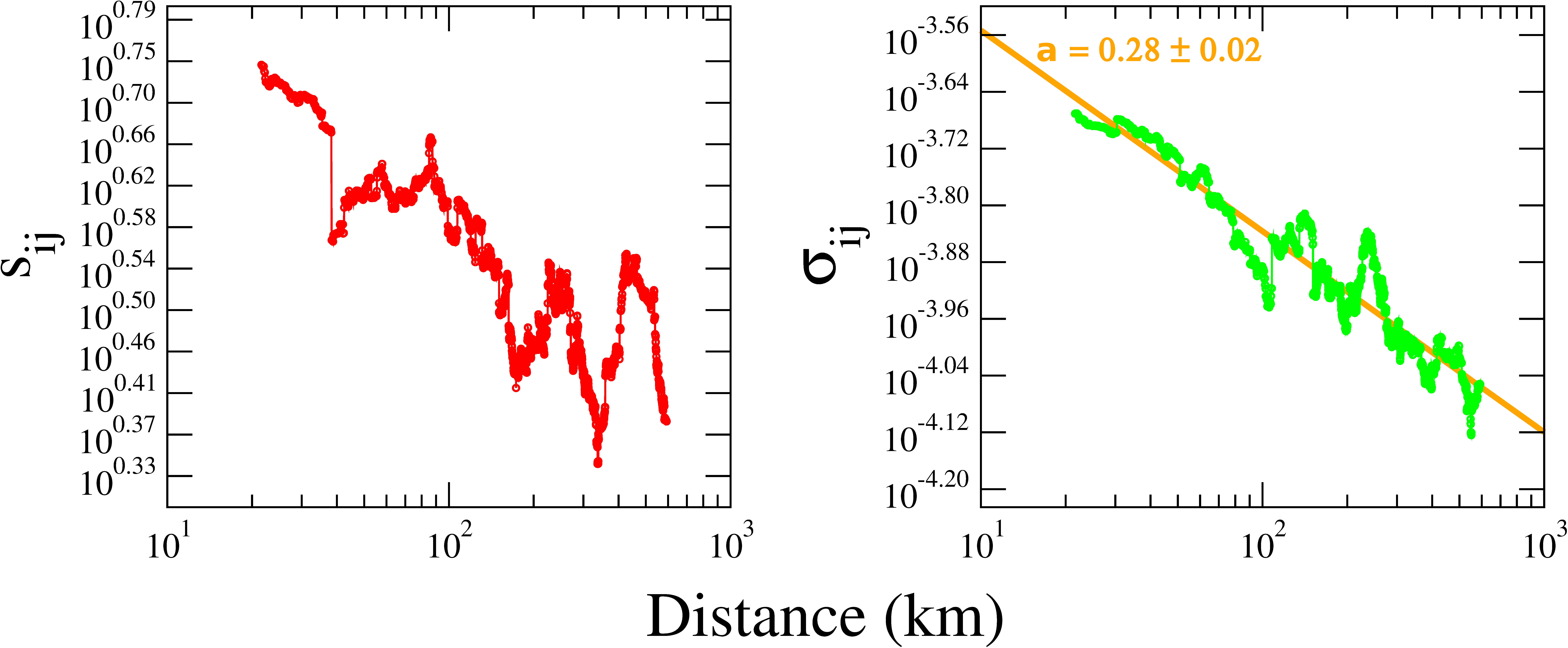}
	\caption{Left: Raw hyperlink strength ($S_{ij}$) between universities versus geographical distance, and Right: Normalized hyperlink
	strength ($\sigma_{ij}=\frac{S_{ij}}{S_i^{out}S_j^{in}}$) between universities versus geographical distance. 
	The normalized measure follows a gravity-law model with an exponent of $a=0.28\pm0.02$.}
	\label{fig:gravity}
	\end{center}
\end{figure*}

The standard, na\"{\i}ve gravity law approach would suggest that the number of hyperlinks, or the strength of the connection, between two 
given universities is inversely proportional to the square of the distance between the two universities. We let $S_{ij}$ denote the strength from 
university $i$ to university $j$. Focusing on the data from 2010, the left frame of Figure \ref{fig:gravity}, shows that the relationship between this measure and the 
geographical distance between the two universities is very noisy. To correct for the different sizes of universities and their different
linking practices (some universities may just link more than others), we normalize these strengths. We divide $S_{ij}$ by the sum of the weights of 
all edges coming from university $i$ ($S_i^{out}$) multiplied by the sum of the weights of all edges linking to university $j$ ($S_j^{in}$). We 
denote this normalized measure $\sigma_{ij}$ and plot it against physical distance in the right frame of Figure \ref{fig:gravity}. With this 
normalization, the relationship between distance the number 
of hyperlinks (strength) between universities is very clear. In both frames, we use a moving average window with a length of 500 data points and therefore a lower bound of 20km
is introduced. An upper bound is induced by considering only the universities within the UK in this study. However, the gravity law holds significantly within a distance range of almost two decades.

Letting $d_{ij}$ denote the geographical distance between two universities, we then seek the exponent $a$, which best fits the observed data following $\sigma_{ij} \propto d_{ij}^{-a} $.
Using the least squares methods, we fit a linear function to the logarithmically transformed data and find $a=0.28\pm0.02$, which closely matches the findings of Pan et al.~\cite{pan2012} for citation and collaboration networks. In that study,
Pan et al.~found an exponent of $a=0.30$ for the citation network before any normalization, while finding an even stronger role for geographical 
distance ($a=0.77$) after applying a similar normalization to the one we apply here.

\begin{figure*}
	\begin{center}
		\includegraphics[width=0.8\textwidth]{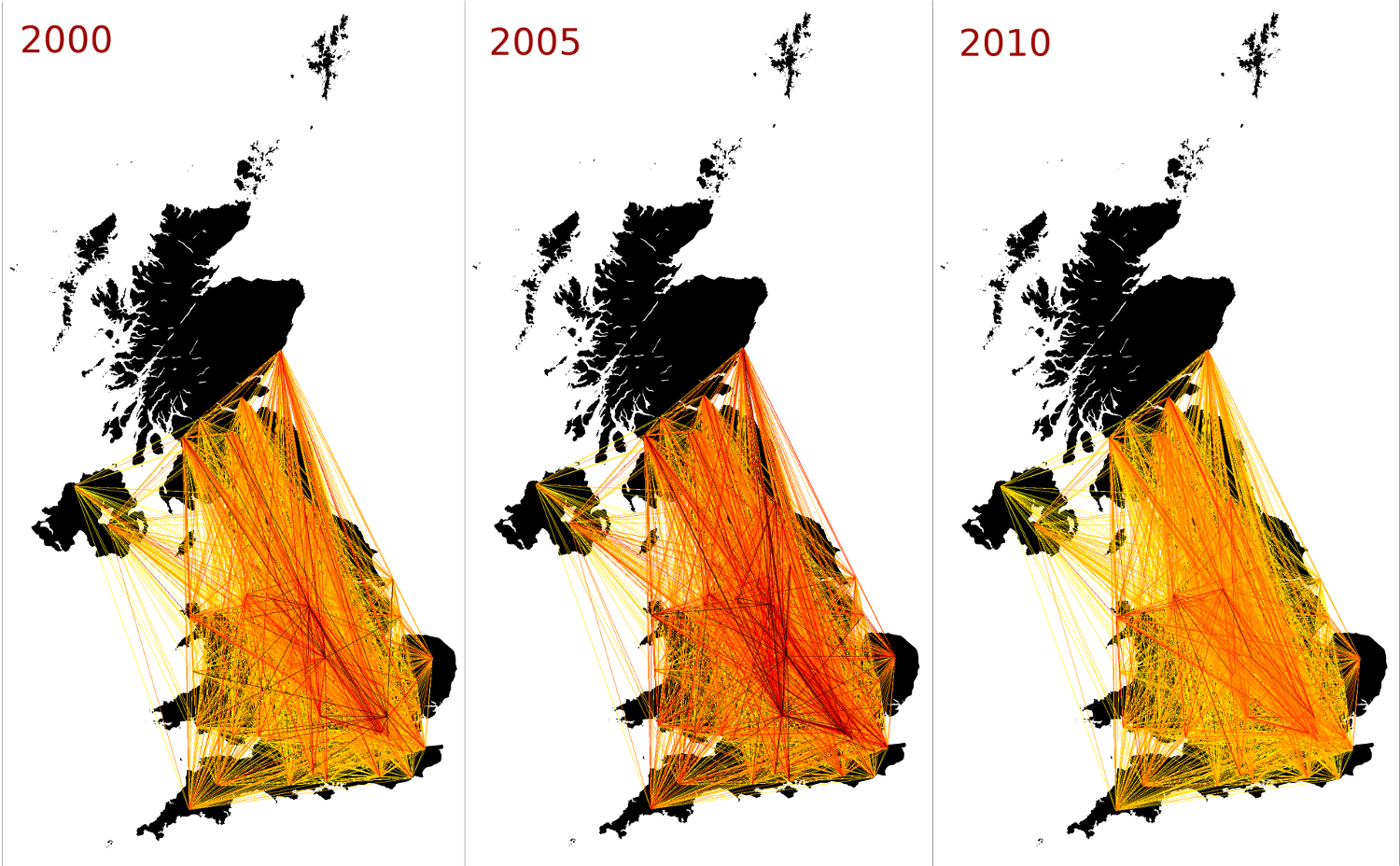}
		\caption{Maps of the UK universities under study for three years: 2000, 2005, and 2010. The connections are the hyperlinks, and color corresponds to the normalized strength of each link ($\sigma_{ij}$). The reddest links correspond to the strongest connections.}
		\label{fig:ac_map}
	\end{center}
\end{figure*}

Figure~\ref{fig:ac_map} maps the universities in the sample along with the connections between them colored according to $\sigma$. It is evident, specially in the map of 2010, that the longer connections generally 
have weaker strength.
It is worth nothing that the size limit of the dataset and the geographical constraints, such as the dense region of London extended to Oxford and Cambridge, which includes a large number of universities in our dataset, could partially drive the strong geographical dependency we observed. This dense region is particularly visible in the map of 2005 in Figure~\ref{fig:ac_map}.

\section{Conclusion}
In this paper, we have reported some of the first findings based on longitudinal analysis of the entire recorded history of the UK web domain.
While this current analysis is by necessity at a macro-level in terms of detail, it nevertheless demonstrates the potential of these
data for detecting changes in patterns in web linking behavior over time, evidence related to the growth and expansion of the web, and uneven 
patterns of linking within subdomains, such as the academic \utt{.ac.uk} subdomain discussed here. We have shown that even though the growth of the 
commercial side of the web has resulted in increasing commercial dominance of the UK web space in terms of absolute number of nodes, the academic 
and government subdomains receive proportionally more inlinks per domain. In examining the academic subdomain in particular, we have shown that while
there is no generalized clustering based on the affiliation of academic institutions, there are clear patterns in terms of higher inlinks to the highest 
status academic institutions and stronger connections between geographically-closer institutions.

This analysis also suggests many future possibilities for research with these web archive data, including more detailed micro-level analysis of 
linking behavior within various subdomains over time, discovery of networks of collaboration between subunits of institutions, comparison between 
link measures and other measures of prominence such as citation networks, and analysis of other subdomains besides \utt{.ac.uk}. In addition, there are 
ongoing efforts to prepare the full-text corpus extracted from the web archive for research (rather than the link corpus used here), which will be 
able to be combined with these data to answer more detailed questions about the content of the web, the context for links, and discourses on the web.

\section{Acknowledgments}
The authors would like to thank Ning Wang for his advice and support on data cleaning and Andreas Kaltenbrunner for his help with creating the
geographic visualizations. The authors are also grateful for funding from UK Jisc for the ``Big Data: Demonstrating the Value 
of the UK Web Domain Dataset for Social Science Research'' grant (16/11 Enhancing the Sustainability of Digital Collections) that 
supported the data extraction and early analysis, and further funding for analysis from the UK Arts and Humanities Research Council 
for the ``Big UK Domain Data for the Arts and Humanities (BUDDAH)'' grant (AH/L009854/1). Finally, the authors would like to thank our anonymous reviewers for their helpful comments on an earlier version of this paper.

\FloatBarrier 
%
\bibliographystyle{abbrv}
\bibliography{library}  

\begin{thebibliography}{10}

\bibitem{ainsworth2011}
S.~G. Ainsworth, A.~Alsum, H.~SalahEldeen, M.~C. Weigle, and M.~L. Nelson.
\newblock How much of the web is archived?
\newblock In {\em Proceedings of the 11th Annual International ACM/IEEE Joint
  Conference on Digital Libraries}, pages 133--136. ACM, 2011.

\bibitem{baeza2007}
R.~Baeza-Yates, C.~Castillo, and E.~N. Efthimiadis.
\newblock Characterization of national web domains.
\newblock {\em ACM Transactions on Internet Technology (TOIT)}, 7(2):9, 2007.

\bibitem{bordino2008}
I.~Bordino, P.~Boldi, D.~Donato, M.~Santini, and S.~Vigna.
\newblock Temporal evolution of the {UK} web.
\newblock In {\em Data Mining Workshops, 2008. ICDMW '08. IEEE International
  Conference on}, pages 909--918, Dec 2008.

\bibitem{brugger2013}
N.~Br{\"u}gger.
\newblock Historical network analysis of the {Web}.
\newblock {\em Social Science Computer Review}, 31(3):306--321, 2013.

\bibitem{brugger2014}
N.~Br{\"u}gger.
\newblock Probing a nation's web sphere: A new approach to web history and a
  new kind of historical source.
\newblock 2014.

\bibitem{dougherty2014}
M.~Dougherty and E.~T. Meyer.
\newblock Community, tools, and practices in web archiving: The state of the
  art in relation to social science and humanities research needs.
\newblock {\em Journal of the American Society of Information Science \&
  Technology}, 2014.

\bibitem{dougherty2010}
M.~Dougherty, E.~T. Meyer, C.~Madsen, C.~V. den Heuvel, A.~Thomas, and
  S.~Wyatt.
\newblock Researcher engagement with web archives: State of the art.
\newblock Technical report, 2010.

\bibitem{escher2006}
T.~Escher, H.~Margetts, V.~Petricek, and I.~Cox.
\newblock Governing from the centre? {C}omparing the nodality of digital
  governments.
\newblock In {\em Annual Meeting of the American Political Science
  Association}, 2006.

\bibitem{foot2006}
K.~A. Foot and S.~M. Schneider.
\newblock {\em Web Campaigning}.
\newblock The MIT Press, 2006.

\bibitem{hale2012}
S.~A. Hale.
\newblock Net increase? {C}ross-lingual linking in the blogosphere.
\newblock {\em Journal of Computer-Mediated Communication}, 17(2):135--151,
  2012.

\bibitem{jisc_intro}
S.~A. Hale, T.~Yasseri, and H.~Margetts.
\newblock Extracting clean hyperlink and website data from the {JISC UK Web
  Domain Dataset}.
\newblock Technical report, 2014.

\bibitem{kahle1997}
B.~Kahle.
\newblock Preserving the {Internet}.
\newblock {\em Scientific American}, 276(3):82--83, 1997.

\bibitem{leiner2009}
B.~M. Leiner, V.~G. Cerf, D.~D. Clark, R.~E. Kahn, L.~Kleinrock, D.~C. Lynch,
  J.~Postel, L.~G. Roberts, and S.~Wolff.
\newblock A brief history of the {Internet}.
\newblock {\em ACM SIGCOMM Computer Communication Review}, 39(5):22--31, 2009.

\bibitem{masanes2006}
J.~Masan{\`e}s.
\newblock Web archiving: Issues and methods.
\newblock In J.~Masan{\`e}s, editor, {\em Web Archiving}, pages 1--54.
  Springer, 2006.

\bibitem{merton1968}
R.~K. Merton.
\newblock The {Matthew} effect in science.
\newblock {\em Science}, 159(3810):56--63, 1968.

\bibitem{newman2006}
M.~E.~J. Newman.
\newblock Modularity and community structure in networks.
\newblock {\em Proceedings of the National Academy of Sciences},
  103(23):8577--8582, 2006.

\bibitem{pan2012}
R.~K. Pan, K.~Kaski, and S.~Fortunato.
\newblock World citation and collaboration networks: Uncovering the role of
  geography in science.
\newblock {\em Scientific Reports}, 2, 2012.

\bibitem{payne2008}
N.~Payne and M.~Thelwall.
\newblock Longitudinal trends in academic {Web} links.
\newblock {\em Journal of Information Science}, 34(1):3--14, 2008.

\bibitem{rogers2000}
R.~Rogers and N.~Marres.
\newblock Landscaping climate change: A mapping technique for understanding
  science and technology debates on the {World Wide Web}.
\newblock {\em Public Understanding of Science}, 9(2):141--163, 2000.

\bibitem{thelwall2003}
M.~Thelwall, R.~Tang, and L.~Price.
\newblock Linguistic patterns of academic {Web} use in {Western} {Europe}.
\newblock {\em Scientometrics}, 56(3):417--432, 2003.

\end{thebibliography}
%
%

\end{document}